\documentclass{article}
\usepackage{graphicx}
\usepackage{amssym}

\newcommand{\rbox}{\begin{flushright}
        \vspace{-8mm}
        \qed
        \vspace{-1mm}
        \end{flushright}
}

\def\If{\underline{If}~}

\def\Else{\underline{Else}~}

\def\While{\underline{While}~}
\def\Loop{\underline{Loop}~}
\def\Until{\underline{Until}~}

\def\eps{\varepsilon}

\def\T{{\cal T}}

\newenvironment{proof}{\noindent{\bf Proof:}}{{\rbox}}
\newenvironment{sketch}{\noindent{\bf Proof Sketch:}}{{\rbox}}
\def\reals{{I\!\!R}}
\def\Reals{{\reals}}

\newtheorem{Claim}{Claim}[section]
\newtheorem{Lemma}[Claim]{Lemma}
\newtheorem{corollary}[Claim]{Corollary}

\renewcommand{\reals}{{\Bbb R}}

\def\epsilon{\varepsilon}

\long\gdef\boxit#1{\vspace{5mm}\begingroup\vbox{\hrule\hbox{\vrule\kern3pt
\vbox{\kern4pt#1\kern3pt}\kern3pt\vrule}\hrule}\endgroup}
\newcommand{\qed}{\vspace{.15cm}\rule{6pt}{6pt}}

\newcommand{\comment}[1]{}

\title{Computing Homotopic Shortest Paths Efficiently
}

\author{
Alon Efrat\\
Department of Computer Science\\
University of Arizona\\
{\tt alon@cs.arizona.edu}
\and
Stephen G.~Kobourov\\
Department of Computer Science\\
University of Arizona\\
{\tt kobourov@cs.arizona.edu}
\and
Anna Lubiw\\
Department of Computer Science\\
University of Waterloo\\
{\tt alubiw@uwaterloo.ca}
\and
}

\begin{document}
\def\polylog{{{\sl polylog }}}

\maketitle
\begin{abstract} \small\baselineskip=9pt

This paper addresses the problem of finding shortest paths homotopic
to a given disjoint set of paths that wind amongst point obstacles in
the plane. We present a faster algorithm than previously known. 

\end{abstract}

\section{Introduction} \label{section:introduction}

Finding Euclidean shortest paths in simple polygons is a well-studied
problem. The funnel algorithm of Chazelle~\cite{FOCS82*339} and Lee
and Preparata~\cite{LP84} finds the shortest path between two points
in a simple polygon. Hershberger and Snoeyink~\cite{HerSno94} unify
earlier results for computing shortest paths in polygons. They
optimize a given path among obstacles in the plane under the Euclidean
and link metrics and under polygonal convex distance functions.
Related work has been done in addressing a classic VLSI problem, the
continuous homotopic routing problem~\cite{cs-rrew-84,lm-artr-85}. For
this problem it is required to route wires with fixed terminals among
fixed obstacles when a sketch of the wires is given, i.e., each wire
is given a specified homotopy class. If the wiring sketch is not given
or the terminals are not fixed, the problem is
NP-hard~\cite{Leiserson:1983:OPR,pinter83river-routing,Richards84b}.

Some of the early work on continuous homotopic routing was done by
Cole and Siegel~\cite{cs-rrew-84} and Leiserson and
Maley~\cite{lm-artr-85}. They show that in the $L_\infty$ norm a solution
can be found in $O(k^3 \log n)$ time and $O(k^3)$ space, where $n$ is
the number of wires and $k$ is the maximum of the input and output
complexities of the wiring. Maley~\cite{maley-thesis} shows how to
extend the distance metric to arbitrary polygonal distance functions
(including Euclidean distance) and presents a $O(k^4 \log n)$ time and
$O(k^4)$ space algorithm.  The best result so far is due to Gao {\em
et al.}~\cite{Storb88} who present a $O(kn^2 \log(kn))$ time and
$O(kn^2)$ space algorithm. Duncan {\em et al.}~\cite{dekw-dwfe01} and Efrat {\rm et al.}~\cite{eksw-gfg-02}
present an $O(kn+n^3)$ algorithm for the related fat edge drawing
problem: given a planar weighted graph $G$ with maximum degree 1 and
an embedding for $G$, find a planar drawing such that all the edges
are drawn as thick as possible and proportional to the corresponding
edge weights.

The topological notion of homotopy formally captures the notion of
deforming paths. Let $\alpha,\beta:[0,1]\longrightarrow\Reals^2$ be two continuous
curves parameterized by arc-length.  Then $\alpha$ and $\beta$ are
{\it homotopic\/}
with respect to a set of obstacles $V\subseteq \reals^2$
if $\alpha$ can be continuously deformed into $\beta$ while avoiding the obstacles;
more formally,
if there exists a continuous
function $h:[0,1]\times [0,1]
\rightarrow \Reals^2$ with the following three properties:
\newpage
\begin{enumerate}
\item $h(0,t)=\alpha(t)$ and $h(1,t)=\beta(t)$, for $0\leq t \leq 1$
\item $h(\lambda,0)=\alpha(0)=\beta(0)$ and $h(\lambda,1)=\alpha(1)=\beta(1)$ for $0\leq
\lambda \leq 1$
\item $h(\lambda,t) \notin V$ for $0 \leq \lambda \leq 1, 0 < t<1$
\end{enumerate}

Let $\Pi=\{\pi_1, \pi_2, \dots, \pi_n\}$ be a set of disjoint, simple
polygonal paths and let the endpoints of the paths in $\Pi$ define the
set $T$ of at most $2n$ fixed points in the plane.
Note that we allow a path to degenerate to a single point.
We call the fixed points of
$T$ ``terminals,'' and call the interior vertices of the paths ``bends,''
and use ``points'' in a more generic sense, e.g.  ``a point in the
plane,'' or ``a point on a path.''
We assume that no two terminals/bends lie on the same vertical line.

\begin{figure}[t]
\begin{center}
{\includegraphics[width=12cm]{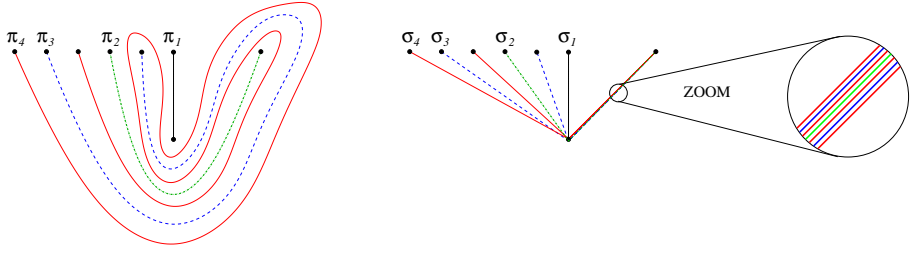}}
\caption{\small\sf
    An example with exponential complexity: $k=\Omega(2^n)$: On the left
    is the initial wiring sketch, and on the right is the wiring after the shortest paths have been computed. The number of edge segments in the shortest paths $\sigma_1, \sigma_2, \sigma_3, \sigma_4$ is $1,2,4,8$, respectively. In general, wire $\sigma_i$ has $2^{i-1}$ edge segments. Note that on the right many edge segments are parallel. }
\label{fig-complexity}
\end{center}
\end{figure}

Our goal is to replace each path $\pi_i\in \Pi$ by a shortest path $\sigma_i$
that
is homotopic to $\pi_i$ with respect to the set of obstacles $T$, see Fig.~\ref{fig-complexity} and Fig.~\ref{fig-monotone}.
Note that $\sigma_i$ is unique.  Let $\Sigma = \{\sigma_1, \ldots
\sigma_n\}$ be the set of resulting paths.
Observe
that these output paths may [self] intersect by way of segments
lying on top of each other, but will be
{\it non-crossing\/} in the sense that a slight perturbation of the bends
makes the paths simple and disjoint.

Let $k_{\it in}$ be the number of edges in all the paths of $\Pi$.
Let $k_{\it out}$ be the number of edges in all the paths of $\Sigma$.  We
will measure the complexity of our algorithms in terms of $n$ and $k =
\max\{k_{\it in}, k_{\it out}\}$.  In fact, the only relationship guaranteed
among the parameters $n, k_{\it in}, k_{\it out}$ is $k_{\it out} \le
nk_{\it in}$. Note that $k$ may be arbitrarily large compared to $n$.
Clearly this is the case since, for example, a path can wind around a
set of terminals arbitrarily many times. However, there exist
non-trivial cases in which $k$ can be much larger than $n$. Even after
shortest paths have been computed for each wire, $k$ can be as large
as $k=\Omega(2^n)$, see Fig.~\ref{fig-complexity}.
The algorithm of Hershberger and
Snoeyink~\cite{HerSno94} runs in time $O(nk)$. The algorithm presented in
this paper runs in time $O(n^{2+\eps} + k \log^2 n)$, which is an
improvement for $n<k$.

Although $k$ can be arbitrarily large compared to $n$, one easily
forms the intuition that, because the paths are simple and disjoint,
$k$ can be large in a non-trivial way only because path sections are
repeated over and over.  For example, a path may spiral arbitrarily
many times around a set of points, but each wrap around the set is the
same.

Our method makes essential use of this observation.  We do not begin
by explicitly searching for repeated path sections---this seems
difficult modulo homotopic equivalence.  Instead we begin in section 2
by applying vertical shortcuts to the paths (homotopically) so
that each left and each right local
extreme point occurs at a terminal.  These terminals must then be part of the
final shortest paths, and we have decomposed the paths into $x$-monotone
pieces with endpoints at terminals.
In section 3 we argue that the number of homotopically
distinct $x$-monotone
pieces is at most $O(n)$.
We will bundle together all the homotopically equivalent pieces.
Routing one representative from each such bundle using the
straightforward ``funnel'' technique takes $O(k + n^2)$ time total.  In
section 3 we reduce this using a ``shielding technique'' where we
again exploit the fact that the paths are disjoint and use the
knowledge gained in routing one shortest path to avoid repeating work
when we route subsequent paths.
The final step of the algorithm is to unbundle, and recover the final
paths by putting together the appropriate pieces.
We summarize the main steps of the algorithm in Fig.~\ref{fig-summary}.

\begin{figure}[t]
\medskip
\noindent
\begin{center}
 \fbox{
 \begin{minipage}{9.5in}
    \baselineskip 1mm
    \begin{tabbing}
       \hspace{6mm}\=\hspace{6mm}\=\hspace{6mm}\=\hspace{5mm}\=\hspace{5mm}\=\hspace{5mm}\=\hspace{5mm}\=\hspace{5mm}\=\hspace{5mm}\=\hspace{5mm}\=\hspace{5mm}\= \kill
{\bf Main Algorithm}\\
\> 1. shortcut paths to divide into monotone pieces\\
\> 2. bundle homotopically equivalent pieces\\
\> 3. find the shortest path for each bundle\\
\> 4. unbundle to recover final paths
       \end{tabbing}
       \end{minipage}
       }
\end{center}
\smallskip
\caption{\small\sf Summary of the algorithm.}
\label{fig-summary}
\end{figure}

Before we turn to the remainder of the paper, which consists of one
section for each of steps 1, 2, and 3, it is worth noting that we rely on
some powerful techniques.  In step 1, we use simplex range search
queries to perform homotopic simplifications on the paths.  To do
this, we use Chazelle's cutting trees data
structure~\cite{Chazelle93b}.  In step 2 we need to identify
homotopically equivalent monotone path pieces.  We use the efficient
trapezoidization algorithm of Bar-Yehuda and
Chazelle~\cite{BC-tdjc-94} to perform this step.  Step 3 uses ideas
from ``funnel'' algorithms for shortest paths~\cite{LP84}.
Step 4 is straight-forward.

We will find it convenient to regard each terminal as a small diamond.
Otherwise when a vertical shortcut goes through a terminal we will
need to specify whether the path actually goes to the left or the right of
the terminal, which is awkward to say and to draw.

\section{Shortcutting to Divide Paths into Monotone Pieces}

\begin{figure}[h]
\begin{center}
{\includegraphics[width=12cm]{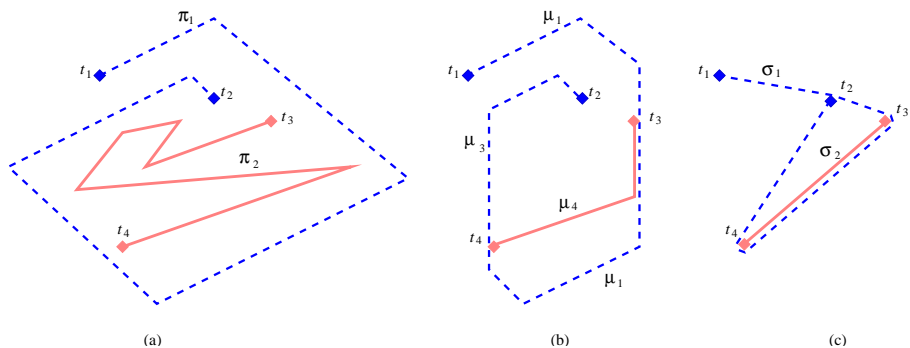}}
\caption{\small\sf
(a)
Two paths $\pi_1$ and $pi_2$ joining terminals $t_1$ to $t_2$ and $t_3$
to $t_4$, respectively. (b) The two paths after performing vertical
shortcuts. Note that path $\pi_1$ now consists of 3 monotone pieces: $\mu_1$ from $t_1$ to $t_3$, $\mu_2$ from $t_3$ to $t_4$, and $\mu_3$ from $t_4$ to $t_2$; path $\pi_2$ now
consists of one $x$-monotone piece, $\mu_4$, homotopically equivalent to $\mu_2$; (c) The final homotopically equivalent
shortest paths, $\sigma_1$ and $\sigma_2$.  }
\label{fig-monotone}
\end{center}
\end{figure}

We begin by applying vertical shortcuts to reduce each path to a
sequence of $x$-monotone path sections, see Fig.~\ref{fig-monotone}.
A {\it vertical
shortcut\/} is a vertical line segment $ab$ joining a point $a$ on
some path $\pi$ with a point $b$ also on $\pi$, and with the property that
the subpath of $\pi$ joining $a$ and $b$, $\pi_{ab}$, is homotopic to the
line segment $ab$.

\begin{figure}[t]
\begin{center}
{\includegraphics[width=14cm]{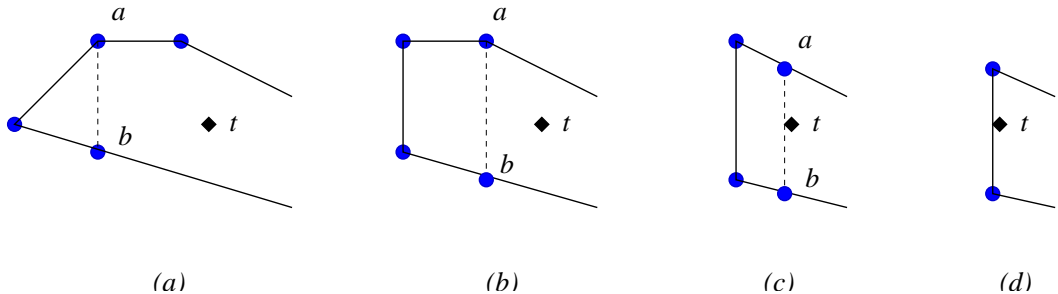}}
 \caption{\small\sf
Maximal left shortcuts (bends are represented by circles and
terminals by diamonds):
(a) a 2 segment shortcut, maximal because $a$ is a bend; (b) a 3 segment
shortcut, maximal for the same reason; (c) a 3 segment shortcut,
maximal because $ab$ hits a terminal; (d) the result of the shortcut in (c).
}
\label{fig-left}
\end{center}
\end{figure}

\begin{figure}[t]
\begin{center}
{\includegraphics[width=14cm]{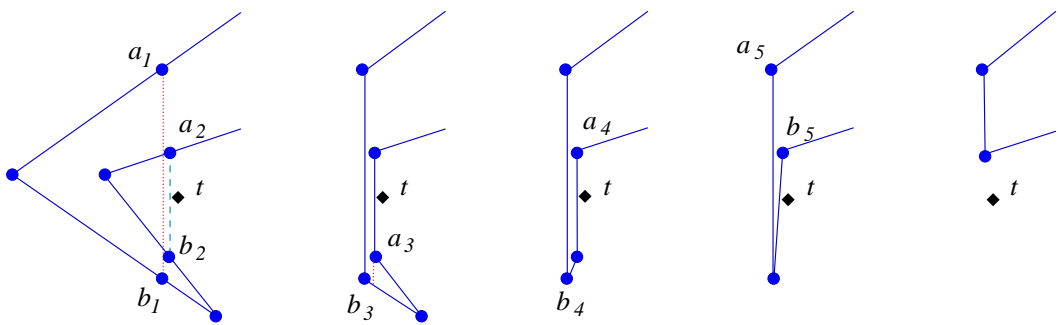}}
 \caption{\small\sf
A sequence of maximal elementary vertical shortcuts
(bends are represented by circles and terminals by diamonds).
Shortcuts $a_1b_1$ and $a_2b_2$ are left shortcuts; $a_3b_3$ is a right
shortcut; and $a_4b_4$ and $a_5b_5$ are collinear shortcuts.
}
\label{fig-shortcuts}
\end{center}
\end{figure}

We will only do {\it elementary vertical shortcuts\/} where the subpath
$\pi_{ab}$ consists of [portions of] 2 line segments or
3 line segments with the middle one
vertical, and the other two non-vertical.
We distinguish {\it left shortcuts\/} which are elementary vertical
shortcuts where $\pi_{ab}$ contains a point to the left of the line through $ab$;
{\it right shortcuts\/} where $\pi_{ab}$ contains a point to the right of
the line through $ab$; and {\it collinear shortcuts\/} where $\pi_{ab}$
lies in the line through $ab$, see Fig.~\ref{fig-left} and
Fig.~\ref{fig-shortcuts}.
Collinear shortcuts are always applied after left/right shortcuts
(and only then), and prevent consecutive vertical segments.

We will in fact only apply {\it maximal\/} elementary vertical shortcuts,
where the
subpath $\pi_{ab}$ cannot be increased.  In particular, this means
that for left and right shortcuts, either $a$ or $b$ is a bend,
or the line segment $ab$ hits a terminal.

A {\it local left [right] extreme\/} of a path is
a point or, more generally, a vertical segment,
where the $x$-coordinate of the path reaches a local min [max].
Observe that every left [right] shortcut ``cuts off'' a local left [right]
extreme.  Conversely, every left [right] local extreme provides a left
[right] shortcut {\it unless\/} the local extreme
{\it is locked at\/} a terminal,
meaning that the left [right] extreme contains the left [right] point of the
terminal's diamond.
Figure~\ref{fig-left} shows 4 cases of left extremes; in (a--c) they provide
left shortcuts, but in case (d) the left extreme is locked at a terminal.

We will use range searching to detect, for a local left extreme, $l$,
the maximal left shortcut that can be performed there.
In particular, given $l$, we first identify the maximal potential shortcut
in the absence of terminals.
To do this, take the segment preceding $l$ and the segment following $l$
along the path.  Take their right endpoints, and let $t$ be the leftmost
of these two points.  Then $t$ determines the maximal potential shortcut
that can be performed at $l$.
This potential shortcut forms either
a triangle (for example, triangle $abl$ in Figure~\ref{fig-left}(a))
or a trapezoid (for example, the trapezoid determined by $a$, $b$, and $l$,
in Figure~\ref{fig-left}(b)).
If a range query tells us that the triangle/trapezoid is free of terminals,
then it forms the true maximal shortcut from $l$.
Otherwise, the range query should tell us the leftmost terminal inside
the triangle/trapezoid, and that determines the maximal left shortcut from $l$.
Note that the ranges we need to
query are triangles with one vertical side, and trapezoids with two
vertical sides.

\begin{figure}[t]
\begin{center}
{\includegraphics[width=10cm]{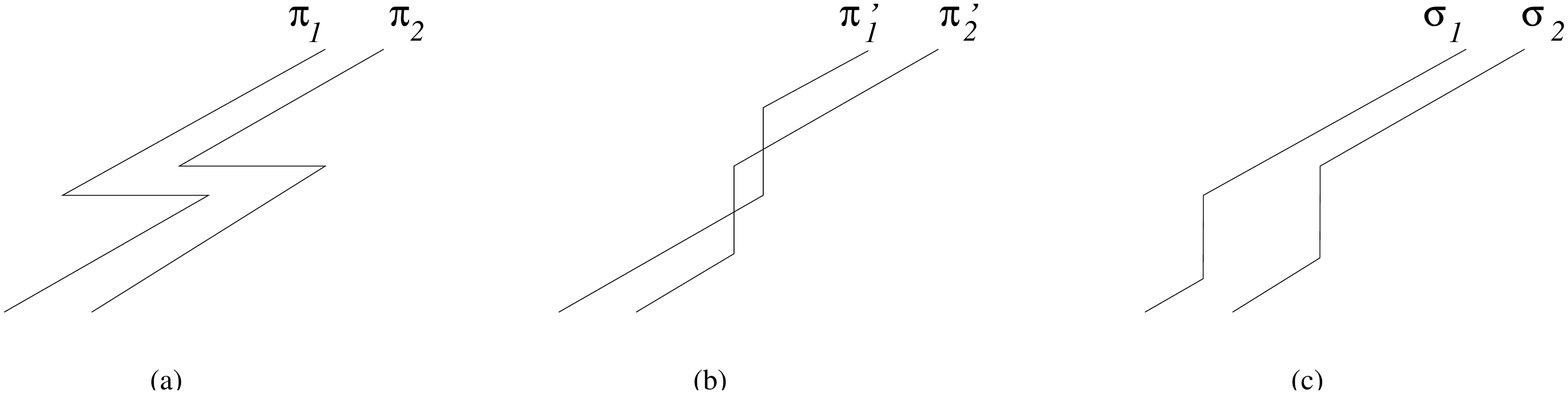}}
 \caption{\small\sf
Performing shortcuts in arbitrary order may result in crossings. (a) Two disjoint paths $w_1$ and $w_2$; (b) After a left shortcut in $w_1$ and a right shortcut in $w_2$ there is a crossing. Note that no further shortcuts are possible for either path, and hence the crossing cannot be removed; (c) If both shortcuts are right [left] there are no crossings.
}
\label{fig-crossing}
\end{center}
\end{figure}

Our algorithm is to perform elementary vertical shortcuts until none
remain, at which time all local left and right extremes are locked at
terminals, so the path divides into monotone pieces.
See Section \ref{section:monotone-correctness} for justification.
Doing elementary vertical shortcuts in an arbitrary order may result in crossing
paths as shown in Fig.~\ref{fig-crossing}; to guarantee
non-crossing paths we will do the elementary vertical shortcuts
in alternating {\it left\/} and {\it right phases\/},
where a {\it left [right] phase\/} means we do
left [right] shortcuts (and the consequent collinear shortcuts)
until no more are possible.
See Section \ref{section:non-crossing} for justification.
We summarize these steps in Fig.~\ref{fig-pseudocode}.

\begin{figure}[thb]
\noindent
 \fbox{
 \begin{minipage}{9.5in}
    \baselineskip 1mm
    \begin{tabbing}
       \hspace{6mm}\=\hspace{6mm}\=\hspace{6mm}\=\hspace{5mm}\=\hspace{5mm}\=\hspace{5mm}\=\hspace{5mm}\=\hspace{5mm}\=\hspace{5mm}\=\hspace{5mm}\=\hspace{5mm}\= \kill

{\tt ShortCut}\\
\vspace{1cm}

traverse paths initializing\\
\> $L$ $\leftarrow$ $\{$local left extremes not locked at terminals$\}$\\
\> $R$ $\leftarrow$ $\{$local right extremes not locked at terminals$\}$\\
\Loop\\
\> {\tt LeftPhase}\\
\> {\tt RightPhase}\\
\Until $L$ and $R$ are empty\\ \\
\vspace{1cm}

{\tt LeftPhase}\\
\While $L$ is non-empty\\
\> \> $l \leftarrow$ remove an element of $L$\\
\> \> $\tau \leftarrow$ the triangle/trapezoid forming the maximal potential shortcut at $l$\\
\> \> RangeQuery($\tau$) \\
\> \>\If $\tau$ is empty of terminals\\
\> \>\>$s \leftarrow$ right side of $\tau$\\
\> \>\Else\\
\> \>\> $s \leftarrow$ leftmost terminal inside $\tau$\\
\> \>do a left shortcut of $\tau$ to $s$\\
\> \>perform consequent collinear shortcuts\\
\> \>\If any right local extremes disappear, remove them from $R$\\
\> \>\If the final vertical segment is a local left [right] extreme and not
locked at a terminal\\
\> \>\>add it to $L$ [$R$, resp.]\\ \\
\vspace{1cm}

{\tt RightPhase}\\
(defined analogous to {\tt LeftShortcut} but with ``left'' and ``right'' exchanged)

       \end{tabbing}
       \end{minipage}
       }
\medskip
\caption{\small\sf Pseudo-code for the shortcutting phase of the algorithm.}
\label{fig-pseudocode}
\end{figure}

The remainder of this section consists of four subsections:
Subsection \ref{section:correctness} justifies organizing the above algorithm
around the sets of left and right extremes, rather than explicitly searching
each time for a shortcut to perform.
Subsection \ref{section:monotone-correctness} establishes the fact that
performing elementary vertical shortcuts enables us to divide the
paths into $x$-monotone pieces with endpoints at terminals.
Subsection \ref{section:non-crossing}
argues that doing elementary vertical shortcuts in left/right phases prevents
crossings.
Finally, Subsection \ref{section:implementation} deals with
implementation and run-time analysis.

\subsection{Correctness} \label{section:correctness}

The point of this section is to show that the above algorithm
correctly maintains the sets $L$ and $R$
of local left and right extremes not locked at terminals.
Consider a left phase.  The set $R$ is explicitly updated whenever it changes.
For the correctness of $L$ we use:

\begin{Claim} \label{claim:left-preservation}
Doing one left shortcut does not alter other local left extremes,
nor the shortcuts that will be performed there.
\end{Claim}

\begin{proof}
A left shortcut only removes left portions of non-vertical segments.
\end{proof}

We note that a similar claim fails for collinear shortcuts: doing one
may prevent others.

\subsection{Correctness of division into monotone pieces} \label{section:monotone-correctness}

\begin{Claim} \label{claim:shortcut-structure}
At the end of a left [right] phase of shortcuts, each local left
[right] extreme is locked at a terminal.
\end{Claim}

\begin{figure}[t]
\begin{center}
{\includegraphics[width=12cm]{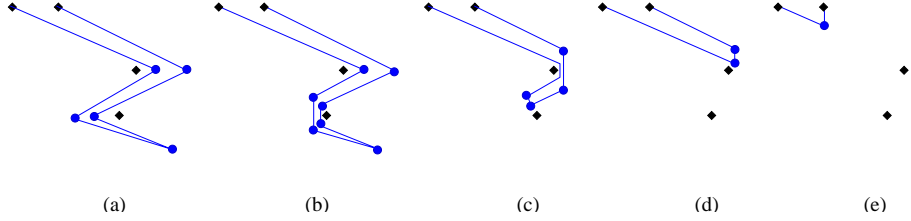}}
\caption{\small\sf
A path going through alternating left and right shortcutting phases: (a) the original path; (b) after a left phase; (c) after a right phase; (d) after a left phase; (e) after a right phase.
 }
\label{fig-many-phases}
\end{center}
\end{figure}

We note that more than one left and one right phase may be required, since
the right phase may add new members to $L$, see Fig.~\ref{fig-many-phases}.

\begin{Claim} \label{claim:shortcut-correctness}
Let $\pi$ be a path, and let
$\mu$ be a result of performing left and right phases
of elementary vertical shortcuts on $\pi$ until no more are possible.
Suppose that the local left and right
extremes of $\mu$ are locked at the terminals $t_{i_1}, \ldots, t_{i_l}$
in that order.
Let $\sigma$ be a shortest path homotopic to $\pi$.
Then the local left [right] extremes of $\sigma$ are locked at exactly the same
ordered list of terminals, and furthermore, the portion of $\sigma$
between $t_{i_j}$ and $t_{i_{j+1}}$ is a shortest path homotopic to the portion
of $\mu$ between those same terminals.

\end{Claim}

\begin{sketch} Because $\pi$ and $\mu$ are homotopic, $\sigma$ is the
shortest path homotopic to $\mu$.  We can thus go from $\mu$ to $\sigma$
using ``rubber band'' deformations that only shorten the path, and such
deformations cannot loosen a left [right] extreme from the terminal
it is locked at.
\end{sketch}

\def\T{{\cal T}}

\subsection{Non-crossing paths}\ \label{section:non-crossing}
The purpose of this section is to prove the following:

\begin{Lemma} Each left [right] phase of shortcuts preserves the
property that paths are non-crossing.
\end{Lemma}

By symmetry, we can concentrate on a left phase.
Note that a left phase, as we have described it, is non-deterministic.
At each stage we choose one member $l$ from
the set $L$ of current local left extremes,
and use it to perform a left shortcut.
To prove the Lemma we will first show that for each left phase there is
{\it some\/} sequence of choices that preserves the property that
the paths are non-crossing.
We will then argue that the end result of a phase does not depend on
the choices made.

\begin{Claim} Suppose we have a set of non-crossing paths.
Let $L$ be the current set of local left extremes not locked at
terminals, and let $l \in L$ be a rightmost element of $L$.
Performing the shortcut for $l$ (together with any consequent collinear
shortcuts) leaves the paths non-crossing.
\end{Claim}

Observe that we can---at least in theory---complete a left phase of shortcuts
using this ``rightmost'' order.
In practice we choose not to
do this simply because of the extra time required to maintain a heap.

\begin{Lemma}
The end result of a left phase does not depend on the sequence of choices
made during the phase.
\end{Lemma}

\begin{proof}
We begin with the claim that the set $\cal L$ of all local left extremes
that appear in $L$ over the course of the phase is independent of the choices
made during the phase.

This implies that the set of left shortcuts performed during the
course of the phase is also independent of the choices made during the phase.
However, the set of collinear shortcuts is {\it not\/} independent.
In particular, the order in which we perform left shortcuts affects
the set of collinear shortcuts, see Fig.~\ref{fig-bigmess}.

\begin{figure}[t]
\begin{center}
{\includegraphics[width=9cm]{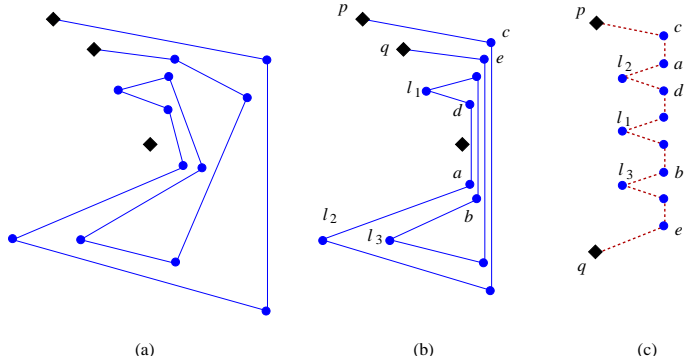}}
\caption{\small\sf
(a) A given path undergoes a right phase; (b)
Handling local left extreme $l_1$ first yields the shortcut $ab$. Handling the local left extreme $l_2$ first yields shortcut $cd$. The final result of handling $l_1$, $l_2$ and $l_3$ in any order is shortcut $ce$; (c) An ``unfolded'' version of part (b).
}
\label{fig-bigmess}
\end{center}
\end{figure}

Consider one left phase.  Let $L_0$ be the initial set of local left
extremes not locked at terminals.  Let $\cal L$ be the union of $L$
over the course of the phase.  Suppose two sequences of choices $C_1$
and $C_2$ during a left phase yield sets ${\cal L}_1$ and ${\cal
L}_2$.  We would like to show that ${\cal L}_1 = {\cal L}_2$.
Consider $l \in {\cal L}_1$.  We will prove $l \in {\cal L}_2$ by
induction on the number of left shortcuts performed in the phase
before $l$ enters ${\cal L}_1$.  If this number is 0 then $l \in L_0$
and we are done.  Otherwise, $l$ is added to $L$ as a result of some
left shortcut and consequent collinear shortcuts.  Any vertical
segment used in the collinear shortcuts may, in its turn, have arisen
as a result of some left shortcut and consequent collinear shortcuts.
Tracing this process, we find that $l$ is formed from a set of left
shortcuts linked by vertical segments, all in the same vertical line
as $l$, see Fig.~\ref{fig-bigmess}.
All these left shortcuts arose from
local left extremes that entered ${\cal L}_1$ before $l$ did, and
thus, by induction, are in ${\cal L}_2$. By
Claim~\ref{claim:left-preservation} left shortcuts are performed at
each of these during the choice sequence $C_2$---though possibly in a
different order than in $C_1$.  The consequent collinear shortcuts
will merge all the verticals forming $l$, and cannot merge more than
that because the segments attached before and after $l$ are not
vertical ($l$ is a local left extreme).  Thus $l$ is in ${\cal L}_2$.

This proves that the set of local left extremes, and thus the set
of left shortcuts is independent of choices made during the phase.
Any vertical segment that is in the final set of paths output
by the phase arises through left shortcuts plus consequent collinear
shortcuts.  Since any set of choices leads to the same set of left shortcuts,
though possibly in different order, the consequent collinear shortcuts
will arrive at the same final vertical shortcuts---i.e. the same
final paths.

\end{proof}

\subsection{Implementation and run-time analysis} \label{section:implementation}

\def\eps{\varepsilon}

In order to perform range queries we need the cutting trees
data structure of Chazelle~\cite{Chazelle93b}. The cutting
trees can be constructed in  $O(n^{2+\eps})$ time and space, where $\eps$ is an
arbitrarily small constant,
and they support simplex range queries in time
$O(\log n)$  ~\cite{Chazelle93b}.
(Note that a trapezoid is a union of two triangles.)
In case a triangle that we query is not empty,
we need to find the rightmost/leftmost terminal inside it.
To do this, create a segment tree $\T$ on the $x$-projections of the terminals,
and maintain a cutting tree on each node of $\T$.
This enables us to find the rightmost/leftmost
terminal inside a query triangle in time $O(\log^2n)$, without
increasing the asymptotic space and preprocessing time.
Details are standard and omitted from this abstract.

\begin{Claim} \label{claim:shortcut-bounds}
The number of elementary vertical shortcuts that
can be applied
to a set of paths with a total of $k$ segments is at most $2k$.
\end{Claim}

\begin{proof}
Assume that no two terminals and/or bends line up vertically.
Consider the set of vertical lines through bends and
through the left and right sides of each terminal's diamond.
An elementary shortcut operates between two of these vertical lines.  If
a left [right] shortcut has its leftmost [rightmost] vertical at a
bend, then after the shortcut, this vertical disappears forever, see Fig.~\ref{fig-shortcuts}.  There are thus at most $k$ such elementary
shortcuts.  Consider, on the other hand, a left shortcut that has its
leftmost vertical at a terminal.  This only occurs when two previous
left shortcuts are stopped at the terminal, and then combined in a collinear
shortcut.  See the right hand pictures of
Fig.~\ref{fig-shortcuts}.  Thus an original edge of the path has disappeared.
Note that elementary shortcuts never fragment an edge of the path into two
edges, but only shorten it from one end or the other.  Thus there are at most
$k$ elementary shortcuts of this type.  Altogether, we obtain a bound
of $2k$ elementary shortcuts.
\end{proof}

This implies the following simple corollary:

\begin{corollary}
The running time of the shortcutting phase, not counting the
preprocessing to construct cutting trees, is $O(k \log^2n )$.
\end{corollary}

Including preprocessing, step 1 takes time
$O(n^{2+\eps} + k \log^2 n)$.

\section{Bundling Homotopically Identical Paths}

Let $M=\{\mu_1\dots \mu_m \}$ be the set of $x$-monotone paths
obtained from step 1 of our algorithm. In the second step of
the algorithm we bundle homotopically equivalent paths in $M$.
More precisely, we take
one representative path for each equivalence class of homotopically
equivalent paths in $M$.
This is justified because the paths in each equivalence class have the
same homotopic shortest path.
Because the paths in $M$ are non-crossing and $x$-monotone, it is
easier to detect
homotopic equivalence: two paths are homotopically equivalent if they
have the same endpoints, and, between these endpoints
no terminal lies vertically above one path and vertically below the other.

\begin{figure}[t]
\begin{center}
{\includegraphics[width=4cm]{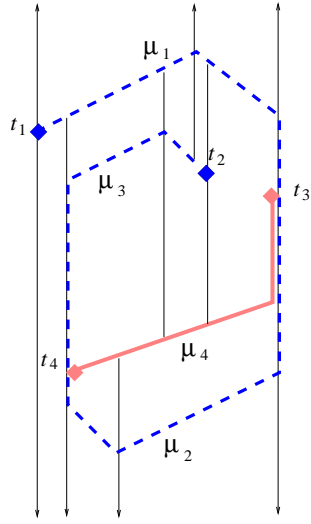}}
\caption{\small\sf
A trapezoidization enables us to identify that $\mu_2$ and $\mu_4$ are homotopically equivalent.
}
\label{fig-trapez}
\end{center}
\end{figure}

In order to perform the bundling we use a
trapezoidization of $M$, see Fig.~\ref{fig-trapez}.  We apply the trapezoidization algorithm of
Bar-Yehuda and Chazelle~\cite{BC-tdjc-94} to the paths obtained after
the shortcuts of step 1, but before these paths are chopped
into monotone pieces.

\begin{Claim} We can perturb the paths output from step 1 so that
they become disjoint and the Bar-Yehuda and Chazelle
algorithm can be applied.
\end{Claim}

The trapezoidization algorithm takes time $O(k + n (\log n)^{1+\eps})$,
which is bounded by the time taken by step 1.
Once we have a trapezoidization of $M$, we bundle homotopically
equivalent paths as follows.  While scanning each $\mu\in M$, we
check if it is homotopically equivalent to the path ``below'' it,
by examining all the edges of the trapezoidization that are incident to
$\mu$ from below.  If all these trapezoidization edges reach the same path
$\mu_j$ and none pass through a terminal on the way to $\mu_j$,
and $\mu_i$ and $\mu_j$ have the same
terminals as endpoints, then we mark
$\mu_i$ as a duplicate. Let $R=\{\rho_1\dots \rho_r \}$
be the paths of $M$ that are not marked as duplicates.

\begin{Lemma}
The number of paths in $R$ is bounded by $2n$.
\end{Lemma}

\begin{proof}
Note that every terminal $t$ is either a right endpoint of paths in $R$
or a left endpoint of paths in $R$, but not both.
This is because a terminal cannot have local left extremes and local 
right extremes locked at it without the paths crossing.

For each terminal $t \in T$, associate its bottommost incident path $\phi(t)$
(its ``floor''), and the first path hit by a vertical ray going up from $t$, 
$\gamma(t)$ (its ``roof'').  
We claim that every path in $R$ is $\phi(t)$ or $\gamma(t)$ for
some $t \in T$.  This proves that the number of paths in $R$ is at most 
$2n$.

Consider a path $\rho \in R$ with left and right terminals $s$ and $t$, 
respectively.   Suppose that $\rho$ is not a roof.  Then every vertical 
ray extending downward from a point of $\rho$ must hit the same path $\sigma$.
(If two rays hit different paths, then in the middle some ray must
hit a terminal.)
Furthermore, since $\rho$ is not the bottommost path incident to $s$,
$\sigma$ must hit $s$.  Similarly $\sigma$ must hit $t$.  But then 
$\sigma$ and $\rho$ are homotopically equivalent.
\end{proof}

\section{Shortest Paths}

In this section we find shortest paths homotopic to the $O(n)$
monotone paths $R = \{\rho_1, \ldots, \rho_r\}$ produced
in the previous section.
Let $\rho'_i$ denote the shortest path homotopic to $\rho_i$.
We route each path using a funnel technique.
The funnel algorithm of
\cite{FOCS82*339} and \cite{LP84}
operates on a triangulation of the $n$ points
(the terminals in our case), and follows the path through the triangulation
maintaining a current ``funnel'' containing all possible shortest paths
to this point.
The algorithm takes time proportional to
the number of edges in the path plus the
number of intersections between
the triangulation edges and the path.

\begin{figure}[t]
\begin{center}
{\includegraphics[width=12cm]{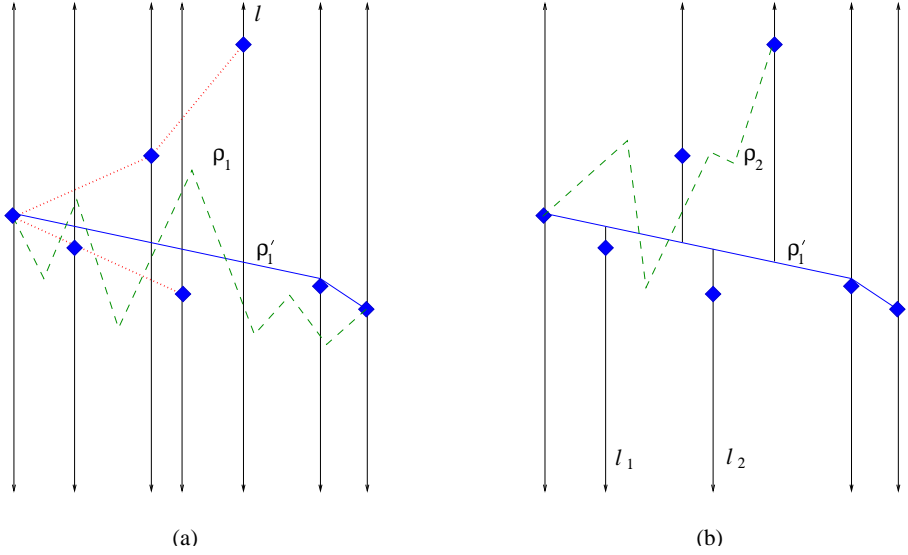}}
\caption{\small\sf
(a) A path $\rho_1$ (dashed), its funnel up to line $l$ (dotted) and the final homotopic shortest path $\rho'_1$; (b) After the shortest path $\rho'_1$ has been found, shielding allows us to route another path $\rho_2$ without examining vertical lines $l_1$ and $l_2$.
}
\label{fig-shield}
\end{center}
\end{figure}

Rather than a triangulation, we will use a trapezoidization formed by
passing a vertical line through each of the $n$ terminals, see Fig.~\ref{fig-shield}(a).
Then, since each path $\rho_i$ is $x$-monotone, it has $O(n)$ intersections with
trapezoid edges, and the funnel algorithm takes
time $O(n + k_i)$, where $k_i$ is the number of edges in $\rho_i$.
This gives a total over all paths of $O(n^2 + k)$.
Recall that $k$ is $\max\{k_{\it in}, k_{\it out}\}$, where
$k_{\it in}$ is the number of edges in all the input paths,
and $k_{\it out}$ is the number of edges in all the output paths.
Note that the number of edges in $R$ is bounded by $k_{\it  in\/}$.

With respect to our whole algorithm, this time of $O(n^2 +k)$ is dominated by
the $O(n^{2+\epsilon})$ time required to create the range query
data structure in step 1.
However, it is interesting to see how much improvement we can make
in this step alone.
In the remainder of this section we describe a randomized algorithm to
route the $O(n)$ monotone paths of $R$ in time $O(n \log n + k)$, and finally
mention a deterministic algorithm with $O(n \sqrt n + k)$ running time.

Both methods use a ``shielding'' technique.  We begin by describing this
idea intuitively.
First note that the $\rho_i$'s can be routed independently, since
none affects the others.
The initial paths $\rho_i$ are non-crossing, and so are the final
shortest paths, $\rho'_i$.  If $\rho_j$ is below $\rho_i$, and we have already
computed
$\rho'_j$, then $\rho'_j$ behaves as a barrier that ``shields''
$\rho_i$ from terminals that are vertically below $\rho'_j$, see Fig.~\ref{fig-shield}(b).

To utilize shielding we will modify the basic trapezoidization described above
as we discover shortest
paths $\rho'_i$.  In particular, the upward vertical ray
through terminal $t$, $u(t)$, will be truncated at the lowest shortest
path $\rho'_i$ that is strictly above $t$ and does not bend at $t$.
The downward vertical ray through terminal $t$, $d(t)$, will be truncated
in an analogous way.

The shortest paths found so far, together with the truncated vertical rays
$u(t)$ and $d(t)$ for each terminal $t$,
partition the plane into
trapezoids, each bounded from left and right by the vertical rays,
and from above and below by shortest paths.
To route a new path $\rho_i$ through this modified trapezoidization
we use the following algorithm.

\vspace{2mm}
\noindent{\bf Routing $\rho_i$ with shielding}
\begin{enumerate}

\item
Identify the first trapezoid that $\rho'_i$ will traverse.  This can
be done in $O(\log n)$ time because the shortest paths observe the
same vertical ordering as the original $\rho_j$'s.

\item
Traverse from left to right the sequence of trapezoids that $\rho'_i$
will pass through.  (Note that $\rho_i$ itself may pass through different
trapezoids, see Fig.~\ref{fig-shield}(b).)
We construct the funnel for $\rho'_i$ as we do this traversal.
Suppose that our path enters trapezoid $\tau$.  To leave $\tau$ on the right
we have two cases.
If the right side of $\tau$ is a point, then it is a
terminal $t$, and we are locked between
two paths that terminate or bend at $t$.  Then the funnel collapses
to this point, and we proceed to the next trapezoid if $\rho_i$ continues.
Otherwise the right side of $\tau$ is a vertical through some terminal
$t$, and (unless $\rho_i$ ends at $t$)
we have a choice of two trapezoids to enter, the upper one with
left side $u(t)$ or the lower one with left side $d(t)$.
We follow path $\rho_i$ until it crosses the infinite vertical
line through $t$.  If it passes above $t$ then we enter the upper
trapezoid, and otherwise we enter the lower trapezoid.  We update the
funnel to include $t$.

\item
When we reach the right endpoint of $\rho_i$, the funnel gives
us the shortest homotopic path $\rho'_i$.

\item We update the trapezoidization as follows.
For each vertical segment $u(t)$ or $d(t)$ that is
intersected by $\rho'_i$
we chop  the segment
at its intersection point with $\rho_i'$,
provided that the intersection point is not $t$
(i.e. that $\rho_i'$ does not bend at or terminate at $t$).

\end{enumerate}

Without yet discussing the order in which we route the paths,
we can say a bit about the timing of this shielding method.
As we traverse a path $\rho_i$ we spend time proportional to the
size of $\rho_i$ plus the number of trapezoids traversed by $\rho'_i$.
When $\rho'_i$ leaves a trapezoid, say at the line segment $u(t)$
above terminal $t$, it may happen that $\rho'_i$ will bend at
$t$ or terminate at $t$.  In this case $t$ is part of the output path
$\rho'_i$, and we can charge the work for this trapezoid to the output
size.  If, on the other hand, $\rho'_i$ does not bend or terminate
at $t$, then it crosses $u(t)$ and we chop $u(t)$ there.
In this case we charge the work for this trapezoid to the chop.
Thus the total time spent by the algorithm is $O(k + C)$
where $C$ is the total number of chops performed at the $n$ verticals.

For shielding to be effective
we need to route paths in an order
that makes $C$ grow more slowly than $n^2$.
We first analyze the randomized algorithm where we
choose the next path to route at random with uniform
probability from the remaining paths.

\begin{Claim}
If paths are routed in random order then
the expected number of times we chop a segment $u(t)$ or $d(t)$ is
$O(\log r)$, and thus $C$ is $O(n \log n)$, and the routing takes time
$O(k + n \log n)$.
\end{Claim}

\begin{proof}
This follows from a  standard backward  analysis.
See for example  \cite{cg-book-00} for other proofs along these lines.
Let $l$ be a vertical line through a point $t$,
and let $u$ denote a ray emerging
vertically from $t$.  Assume that $m$ paths
intersecting $u$ have been inserted up to now, and we
are about to insert a new one.  Then $u$ will be chopped if and only if
the new path creates an intersection point below all existing
intersection points on $u$, but above $t$ itself.
Since the order of the
insertions is random, the probability of the new intersection point being
below all other intersection points is $1/m$. Summing
over all paths yields the claimed bound.
\end{proof}

Finally, we mention that we can achieve a routing time of $O(k + n \sqrt n)$
deterministically.  Suppose that the paths $\{\rho_1, \ldots, \rho_r\}$
are in order from top to bottom.  (We get this for free from step 2 of our algorithm.)  Partition the paths into blocks $B_1, \ldots, B_{\sqrt r}$ each
of size $\sqrt r$, and route them one block at a time from $B_{\sqrt r}$ to $B_1$. Within each block we process the paths in order. 
Since the largest increasing [decreasing] sequence with this ordering
has size $\sqrt r$, therefore the number of chops at each vertical
is $\sqrt r$.  Thus $C$ is $O(n \sqrt n)$, and this routing method takes
time $O(k + n \sqrt n)$.

\section{Conclusion and Open Problems}

For any set of $n$ disjoint paths joining pairs of terminals in the plane
we can find shortest paths homotopic 
with respect to the set of terminals
in time $O(n^{2+\eps} + k\log^2 n)$ where $k$ is the sum of input and output 
sizes of the paths.  
If $k$ is larger than $n$ this is better than the 
Hershberger-Snoeyink algorithm which runs in time $O(nk)$.

More generally, we can use any other range search method, and 
obtain a running time of $O(P + kQ + n (\log n)^{1+\eps} + T)$
where $P$ is the preprocessing time for the range search data structure
on $n$ points, $Q$ is the time for a simplex optimization query (find the 
minimum/maximum $x$-coordinate point in a triangle), and $T$ is the 
time for the shortest path method of step 3---$O(n \log n)$ randomized,
or $O(n \sqrt n)$ deterministic.
Whether $k$ is large or small compared to $n$ determines which 
trade-off between $P$ and $Q$ is preferred.
For example, the partition tree method of 
Matou{\v s}ek \cite{Mat93} yields an emptiness 
query time of $O(\sqrt n)$ with near-linear time preprocessing,
and would be preferable for $k=O(n^{3/2})$.
With any approach to range searching, it may be possible to improve
the query time by taking into account the fact that one of the edges
of each query triangle is vertical.

Another open problem is whether we can do without range queries, and
somehow use a trapezoidization of the original paths, since this can
be found so efficiently with the Bar-Yehuda and Chazelle algorithm.

{\small
\bibliographystyle{abbrv}
\bibliography{stephen}
}
\end{document}